\documentclass{b98proc}

\def\Journal#1#2#3#4{{#1} {\bf #2}, #3 (#4)}

\def\NPB{{\em Nucl. Phys.} B}

\def\PLB{{\em Phys. Lett.} B}

\def\PRD{{\em Phys. Rev.} D}

\newcommand{\gsim}{\buildrel > \over {_\sim}}

\begin{document}

\title{
HIGHER-TWIST CONTRIBUTIONS IN EXCLUSIVE PROCESSES
}

\author{M. V\"ANTTINEN$^1$, L. MANKIEWICZ$^{1,2}$, E. STEIN$^3$}
\address{$^1$Technical University of Munich,
Department of Physics, Garching, Germany \\
$^2$N. Copernicus Astronomical Center, Polish Academy of Science, Warsaw \\
$^3$INFN, Sezione di Torino, Italy}


\maketitle

\abstracts{
Using the renormalon technique, we estimate higher-twist contributions
in deeply virtual Compton scattering and in hard exclusive $\pi^0$
leptoproduction.
}

\section{Introduction\label{section:intro}}

The present work deals with exclusive virtual-photon--nucleon processes such
as deeply virtual Compton scattering $\gamma^* N \rightarrow \gamma N'$
(DVCS) and hard exclusive pion production $\gamma^* N \rightarrow \pi N'$.
For kinematical reasons, exclusive processes  always involve a finite
longitudinal momentum transfer and therefore probe matrix
elements of QCD operators between nucleon states of different momenta.
At leading twist ($\tau=2$) and in light-cone gauge, one parton is extracted
from the initial nucleon and another returned to the final nucleon,
carrying a different momentum fraction. These matrix elements can be
expressed in terms of so-called skewed parton
distributions\cite{Ji,Radyushkin} (SPDs), which are functions
of two momentum-fraction variables. SPDs are a generalization of the
more familiar forward parton distribution functions and form factors.

The photon-nucleon scattering amplitude is a convolution of
an SPD and a photon-parton amplitude. In the present work we have
estimated higher-twist corrections, i.e.\ terms suppressed by
$\Lambda^2/Q^2$ where $Q^2$ is the photon virtuality and $\Lambda$
is a hadronic mass scale, to DVCS and pion production. We used
the renormalon\cite{renorm1,renorm2,Beneke} technique
to estimate power-suppressed contributions to the photon-parton
amplitude, which is then convoluted with skewed quark distributions
(unpolarized distributions in DVCS and polarized in pion production).
Working in Radyushkin's formalism\cite{Radyushkin} of double
distributions $F(x,y,\mu^2)$, we employed the model\cite{Radyushkin-model}
\begin{equation}
  F_q(x,y,\mu^2) = h(x,y) (1-x)^{-3} q(x,\mu^2),
\end{equation}
where $q(x,\mu^2)$ is a forward quark distribution and $h(x,y) = 6y(1-x-y)$
("polynomial model") or $h(x,y) = \delta(y-(1-x)/2)$ ("delta-function
model").

\section{The renormalon technique\label{section:renormalons}}

It is well known that the perturbation expansion in gauge theories
diverges for any nonzero coupling constant. Because the expansion
is nevertheless succesful in practice, it is believed to be an
asymptotic expansion.
For an observable $R(\alpha)$ having a perturbation series
$\sum_n r_n \alpha^{n+1}$, the concepts of Borel transformation
$B[R](t) = \sum r_n t^n/n!$ and Borel integral  
\begin{equation}
  \tilde{R}(\alpha) = \int_0^\infty dt \; e^{-t/\alpha} B[R](t)
\end{equation}  
provide a means of assigning a potentially finite number
("sum") $\tilde{R}(\alpha)$ to the divergent series. However,
the Borel integral for QCD observables is not finite because
$B[R]$ has singularities, known as (infrared) renormalon poles,
on the contour of integration. Regularization by deformation of
the contour introduces an ambiguity depending on whether the contour
is chosen below or above the poles. In phenomenological applications
of renormalons, this ambiguity, i.e.\ the residue
of the pole, is used as an estimate of terms beyond the perturbation
series which should be included in the recipe for calculating
$R$. The residues are proportional to $\exp(-t_i/\alpha_s)$,
where $t_i$ is the location of the $i$'th pole. In QCD, where
$1/\alpha_s = \beta_0 \ln(Q^2/\Lambda^2)$, the new terms are thus
suppressed by $(\Lambda^2/Q^2)^{\beta_0 t_i}$.  

A particular source of the renormalon divergence is 
the all-order resummation of loop diagrams with chains of
vacuum-polarization bubbles. In the following we shall calculate
contributions to exclusive amplitudes from diagrams with bubble
chains. After summing over the number of bubbles and Borel transforming,
we evaluate the residue of the pole closest to $t=0$, which will
serve as an estimate of the next-to-leading-twist
contribution to the amplitude. It turns out to be suppressed by
$\Lambda^2/Q^2$, i.e.\ it is of $\tau=4$.

\section{Results}

\begin{figure}[ht]
\centerline{
\resizebox{\textwidth}{!}{
\epsfig{figure=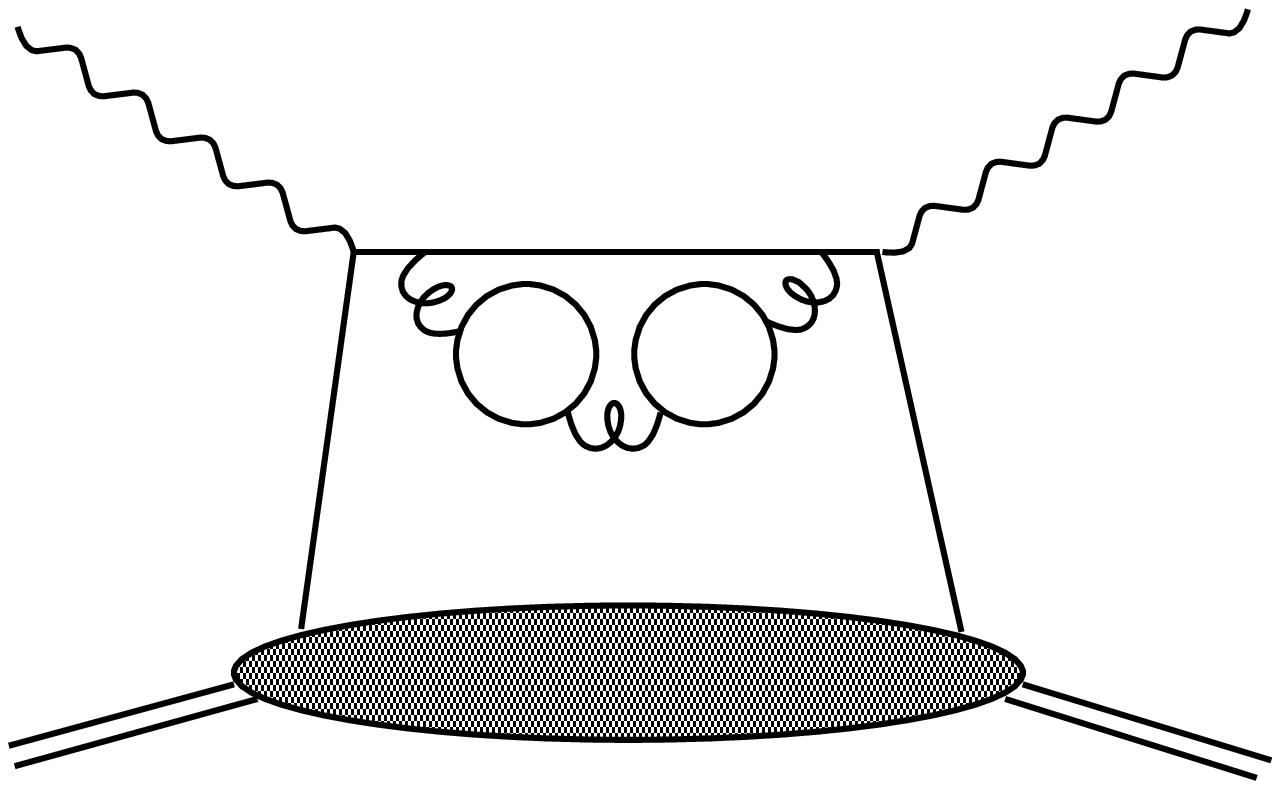}
\epsfig{figure=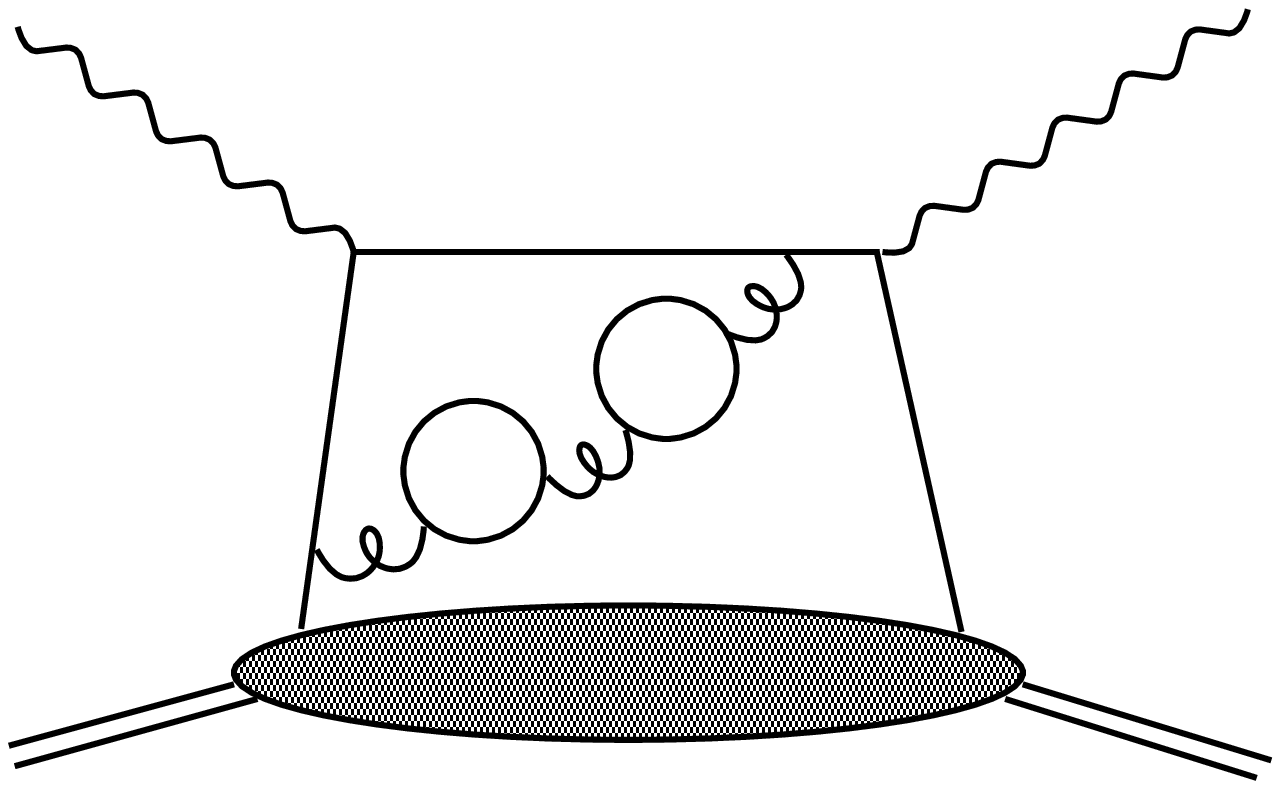}
\epsfig{figure=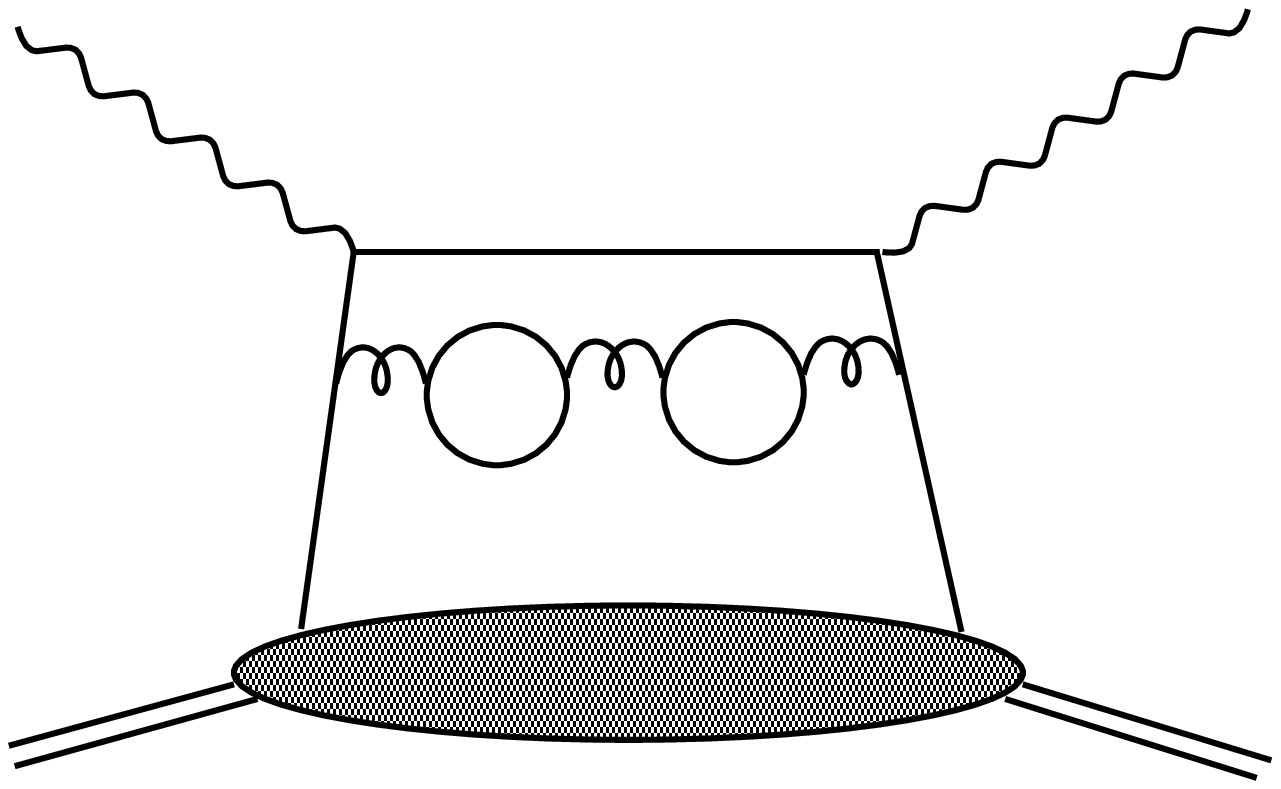}
}}
\centerline{
\resizebox{\textwidth}{!}{
\hspace{1cm}
\epsfig{figure=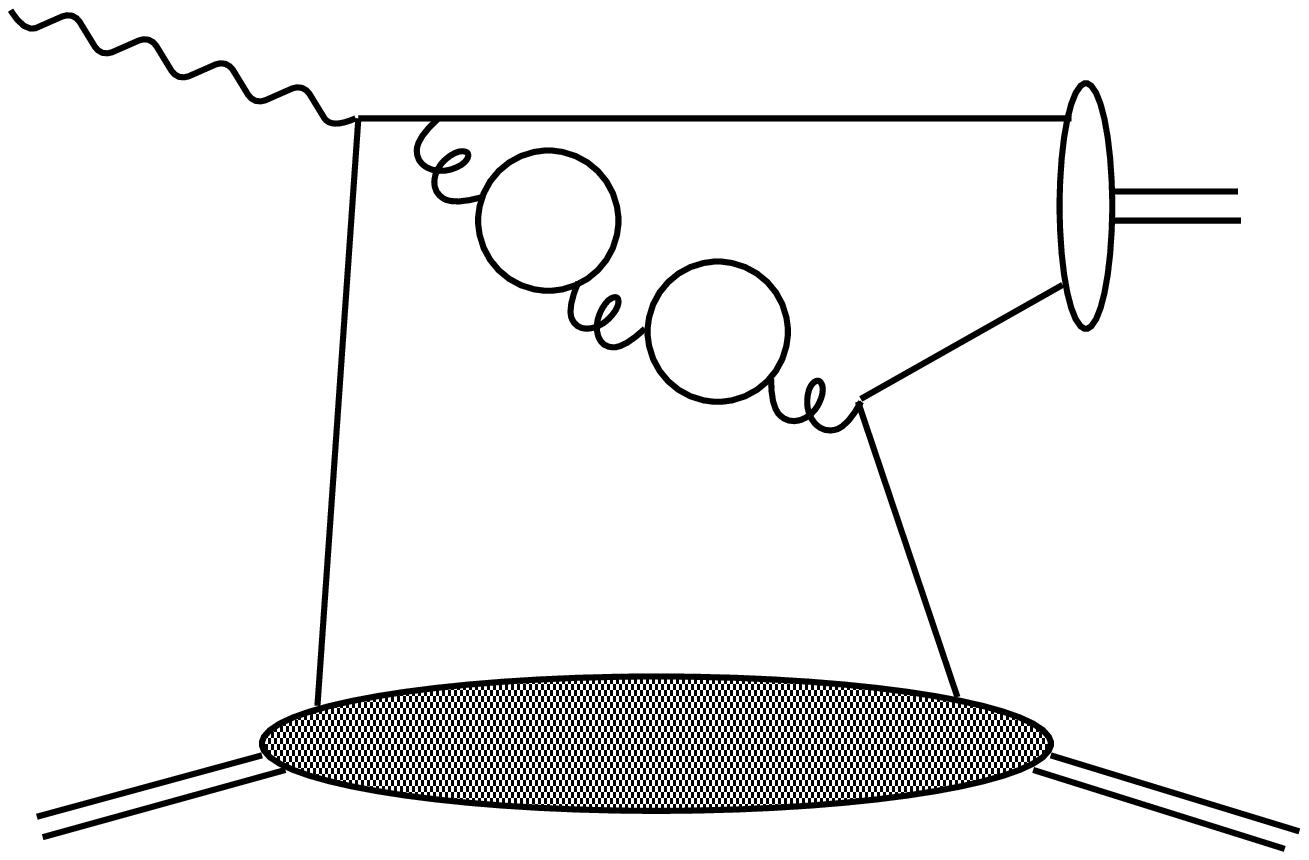,width=2cm}
\epsfig{figure=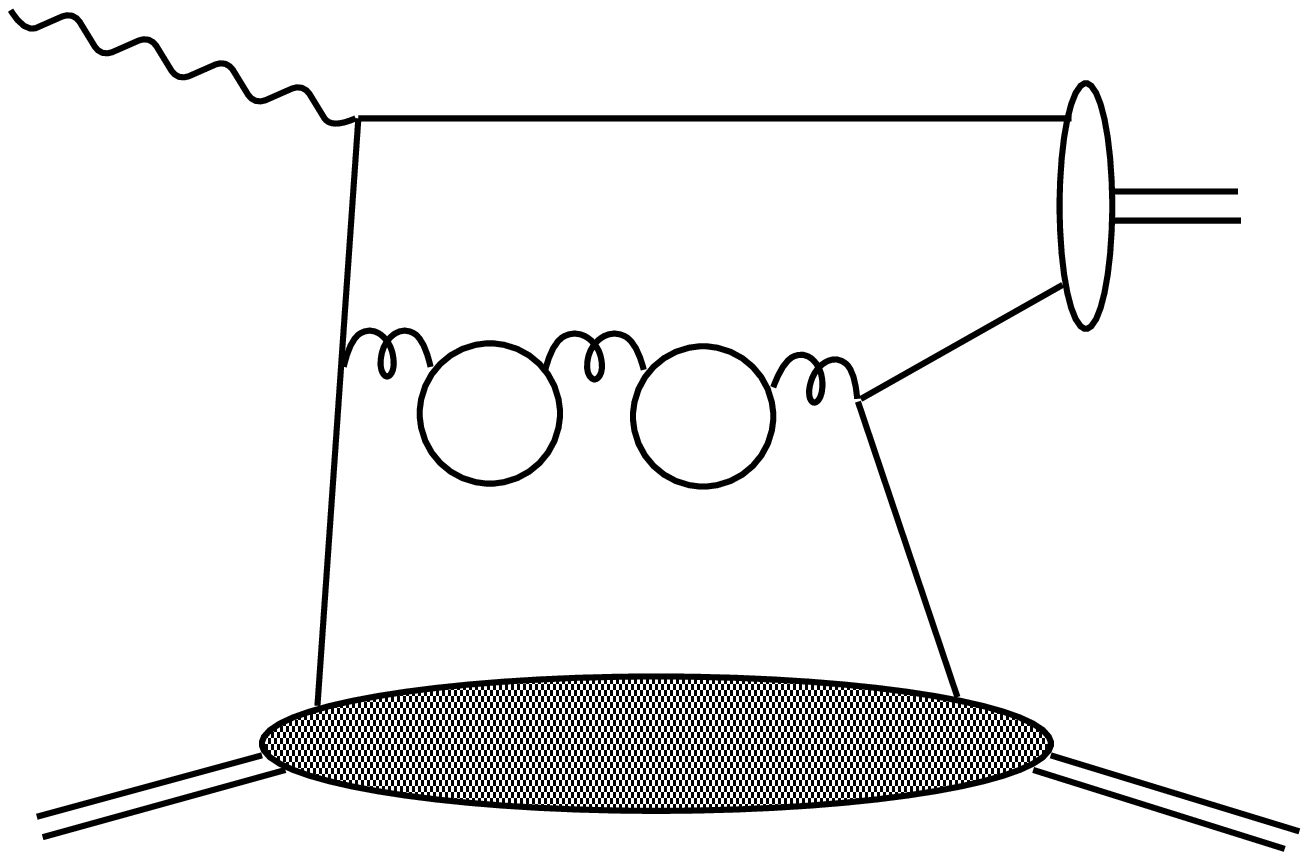,width=2cm}
\hspace{1cm}
}}
\caption{Examples of bubble-chain diagrams contributing to DVCS
(top row) and to hard exclusive meson production (bottom row).} 
\label{fig:diagrams}
\end{figure}

Bubble-chain diagrams giving renormalon contributions to DVCS
and pion production\cite{others} are shown in fig.~\ref{fig:diagrams}.
The leading-twist diagrams for DVCS have no gluons, those for pion
production have a single gluon instead of the bubble chain.  

\begin{figure}[ht]
\resizebox{\textwidth}{!}{\input{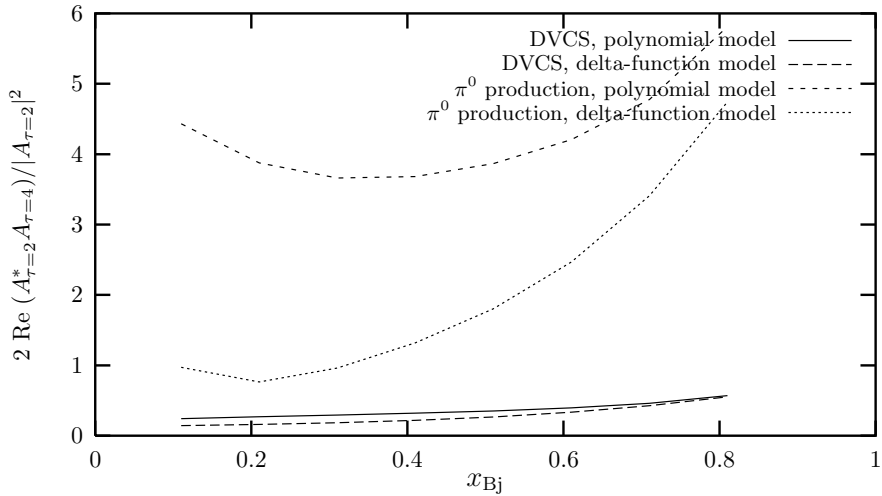}}
\caption{The ratio of estimated higher-twist correction and
leading-twist contribution in DVCS and pion production at $Q^2=4$ GeV$^2$
for two models of SPDs, plotted as a function of Bjorken $x$.}
\label{fig:plot}
\end{figure}

Evaluating the bubble-chain diagrams according to Feynman rules,
taking the residue of the first pole and 
convoluting with model SPDs, we obtain an estimate of the
next-to-leading-twist term in the amplitude $A$. We then calculate
the ratio of the first correction in $|A|^2$ (i.e.\ the interference
of $\tau=2$ and $\tau=4$ amplitudes) and the leading term (square of
$\tau=2$ amplitude), which is plotted in fig.~\ref{fig:plot}
for DVCS and pion production using the two models of SPDs
introduced in section~\ref{section:intro}.

The correction increases with $x_{\rm Bj}$, similarly to renormalon
contributions in inclusive deep inelastic scattering\cite{DIS}. The
magnitude is rather large, especially in pion production, and in fact
we chose $\Lambda^2 = $ (200 MeV)$^2$, i.e.\ a factor 3 smaller than
the value used in fits of inclusive processes. This is not a problem
because theory does not require the normalization factor to be
process-independent. In pion production there is a significant 
difference between the polynomial and delta-function models of SPDs,
whereas in DVCS the two models give similar estimates.

In conclusion, we have found that higher-twist contributions to
exclusive virtual-photon--nucleon processes can be significant at
scales as high as
$Q^2 \simeq 4$ GeV$^2$. We note that at $x \gsim 0.3$ a theoretical
uncertainty arises from the $t$ dependence of the SPDs which we
have not taken into account.

\medskip

{\em Acknowledgements.}
This work was supported in part by BMBF.
M.V. is supported by an Alexander von Humboldt fellowship.
E.S. has been supported by a DFG fellowship.
We wish to thank V.~Braun for illuminating discussions.


\end{document}